\documentclass[a4paper,twocolumn]{article}

\def\FIGUREPATH{./figure/}
\def\RESOURCEPATH{./}

\usepackage{\RESOURCEPATH spconf}
\usepackage{amsmath,amssymb,amsthm}
\usepackage{booktabs, tabularx}
\usepackage{graphicx,subfigure}
\usepackage{epstopdf,bm}
\usepackage{cite}

\ninept

\begin{document}

\title{WAVELET SPEECH ENHANCEMENT BASED ON NONNEGATIVE MATRIX FACTORIZATION}
\name{Syu-Siang Wang$^1$, Alan Chern$^2$, Yu Tsao$^2$, Jeih-weih Hung$^4$, Xugang Lu$^3$, Ying-Hui Lai$^2$, Borching Su$^1$}
\address{$^1$Graduate Institute of Communication Engineering, National Taiwan University, Taiwan\\$^2$Research Center for Information Technology Innovation, Academia Sinica, Taiwan\\$^3$National Institute of Information and Communications Technology, Japan\\$^4$Dept of Electrical Engineering, National Chi Nan University, Taiwan}

\maketitle
\begin{abstract}
For most of the state-of-the-art speech enhancement techniques, a spectrogram is usually preferred than the respective time-domain raw data since it reveals more compact presentation together with conspicuous temporal information over a long time span. However, the short-time Fourier transform (STFT) that creates the spectrogram in general distorts the original signal and thereby limits the capability of the associated speech enhancement techniques. In this study, we propose a novel speech enhancement method that adopts the algorithms of discrete wavelet packet transform (DWPT) and nonnegative matrix factorization (NMF) in order to conquer the aforementioned limitation. In brief, the DWPT is first applied to split a time-domain speech signal into a series of subband signals without introducing any distortion. Then we exploit NMF to highlight the speech component for each subband. Finally, the enhanced subband signals are joined together via the inverse DWPT to reconstruct a noise-reduced signal in time domain.  We evaluate the proposed DWPT-NMF based speech enhancement method on the MHINT task. Experimental results show that this new method behaves very well in prompting speech quality and intelligibility and it outperforms the convnenitional STFT-NMF based method.
\end{abstract}
\section{INTRODUCTION}\label{sec:introduction}
Speech enhancement (SE) techniques, which extract the clean speech component in noise-corrupted utterances, are employed at various applications including hearing aids, speech/speaker recognition, and speech communication over cellphone and internet, to name a few. Increased quality and intelligibility of speech signals benefit the speech interaction in the presence of noise \cite{hartmann2013direct, stark2011use, doclo2009reduced}.

By and large, a conventional spectrum-wise speech enhancement framework \cite{benesty2005speech} consists of two stages, viz. noise tracking and signal gain estimation. The noise tracking stage estimates the power of background noise from the noise-corrupted spectrogram, and some well-known noise tracking schemes include minimum statistics (MS) \cite{martin2001noise}, minima controlled recursive averaging (MCRA) \cite{cohen2001speech}, and improved minima controlled recursive averaging (IMCRA) \cite{cohen2004speech}. As for the stage of signal gain estimation, the noise power information obtained in the first stage is utilized to determine the gain factor on the noise-corrupted spectrogram to predict the embedded clean speech component. The conventional notable methods relative to this stage include spectral subtraction (SS) \cite{boll1979suppression}, Wiener filtering \cite{scalart1996speech}, minimum mean-square error log-spectral amplitude estimation (LSA) \cite{ephraim1985speech,malah1999tracking}, maximum likelihood spectral amplitude estimation (MLSA) \cite{mcaulay1980speech}, maximum a posteriori spectral amplitude estimation (MAPA) \cite{lotter2005speech} and generalized maximum a posteriori spectral amplitude estimation (GMAPA) \cite{conficasspSuTWJ13}. \newline
\indent In particular, the algorithm of nonnegative matrix factorization (NMF) has become one of  the most prevailing data analysis technologies in recent decades, and the speech enhancement methods based on the NMF concept have been well developed and shown very promising results \cite{mohammadiha2013supervised,wilson2008speech}. Briefly speaking, NMF approximates a  data matrix $\mathbf{V}$ by the product of another two matrices, the basis matrix $\mathbf{W}$ and the encoding matrix $\mathbf{H}$, viz. $\mathbf{V}\approx \mathbf{WH}$, where all of the three matrices $\mathbf{V}$, $\mathbf{W}$ and $\mathbf{H}$ contain nonnegative entries only. In NMF-based speech enhancement methods the data matrix $\mathbf{V}$ to be processed is usually the magnitude spectrogram of a speech utterance. Given a pre-trained and fixed basis spectral matrix $\mathbf{W}$ that consists of speech and noise portions, the magnitude spectrogram $\mathbf{V}$ is factorized via NMF.  The resulting approximation $\mathbf{WH}$ for $\mathbf{V}$ is further split into speech and noise components, both of which serve for the signal gain estimation stage of the SE method.  
\newline
\indent The spectrogram being analyzed in the SE methods is mostly created by the well-known short-time Fourier transform (STFT). Despite the success of SE, the STFT operation brings about moderate distortion to the time-domain signal primarily due to the necessary segmentation and windowing processes.  This distortion can be clearly observed when comparing the original signal and the signal reconstructed by the inverse STFT.  However, analyzing the signal directly in time domain is tedious and ineffective since the time-domain signal is usually erratic in waveform and huge in amount.  In light of the aforementioned observations, in this paper we propose to deal with the signal in the wavelet domain for enhancement. In the proposed SE method, a discrete wavelet packet transform (DWPT) is first employed to decompose the time-domain signal into various subband signals. Squaring and framing processes are then applied to each subband signal. Next,  an NMF-wise noise tracking scheme is used to estimate noise with respect to each subband for the subsequent signal gain estimation. Finally,  all of the subband signals are updated via the individual gain and then passed through the inverse DWPT to produce the enhanced time-domain signal.  We evaluate this DWPT-NMF based SE method on the MHINT tasks, and the experimental results confirm that this novel method outperforms the conventional STFT-based SE method and significantly promotes the speech quality and intelligibility under noise-corrupted situations.
\vspace{-0.4cm}
\section{DISCRETE WAVELET PACKET TRANSFORM}\label{sec:dwpt}
The DWPT and inverse DWPT (IDWPT) are performed by a set of well-defined low-pass/high-pass filters together with a down/up-sampling process, and they serve as a perfect (distortionless) analysis/synthesis for an arbitrary signal $\mathbf{f}$. Take DWPT and IDWPT with level 2 for an example. A (full-band) time signal $\mathbf{f}$ is decomposed into two subband signals respectively carrying information of the low- and high-frequency components.  The length of each subband signal is half of that of the original full-band signal due to the factor-2 down-sampling operation.  The decomposition operation is then applied again to each of the two subband signals, and four subband signals are generated accordingly.  The DWPT process can be formulated in Eq. (\ref{eq_dwpt}):

\begin{equation}\label{eq_dwpt}
\mathbf{s}_b^J=DWPT_b^J\{\mathbf{f}\},\quad b=1,2, 3, \cdots,2^J,
\end{equation}
where $\mathbf{s}_b^J$ denotes any subband signal produced by DWPT,  $J$ denotes the level of DWPT, and $b$ refers to the subband index.

The IDWT integrates the subband signals and reconstructs a full-band time signal $\tilde{\mathbf{f}}$. A factor-2 up-sampling operation followed by the pre-defined lowpass/high-pass filters are applied to the four level-2 subband signals, and the resulting two level-1 subband signals undergo the same up-sampling and filtering process to generate a full-band time signal $\tilde{\mathbf{f}}$. Thus the IDWPT can be formulated in Eq. (\ref{eq_idwpt}):
\begin{equation}\label{eq_idwpt}
\tilde{\mathbf{f}}=IDWPT^J\{\mathbf{s}\},
\end{equation}
where $\mathbf{s}$ denotes the set of all level-$J$ subband signals $\{\mathbf{s}_b^J\}_{b=1}^{2^J}$. The reconstructed signal $\tilde{\mathbf{f}}$ will be identical to  the input signal $\mathbf{f}$, namely
\begin{equation}
\tilde{\mathbf{f}}=IDWPT^J\{\mathbf{s}\}=\mathbf{f}.
\end{equation}
\section{STFT-NMF SPEECH ENHANCEMENT SYSTEM}\label{sec:STFTNMFSE}
\begin{figure}[!th]
\centering
\includegraphics[width=.8\columnwidth] {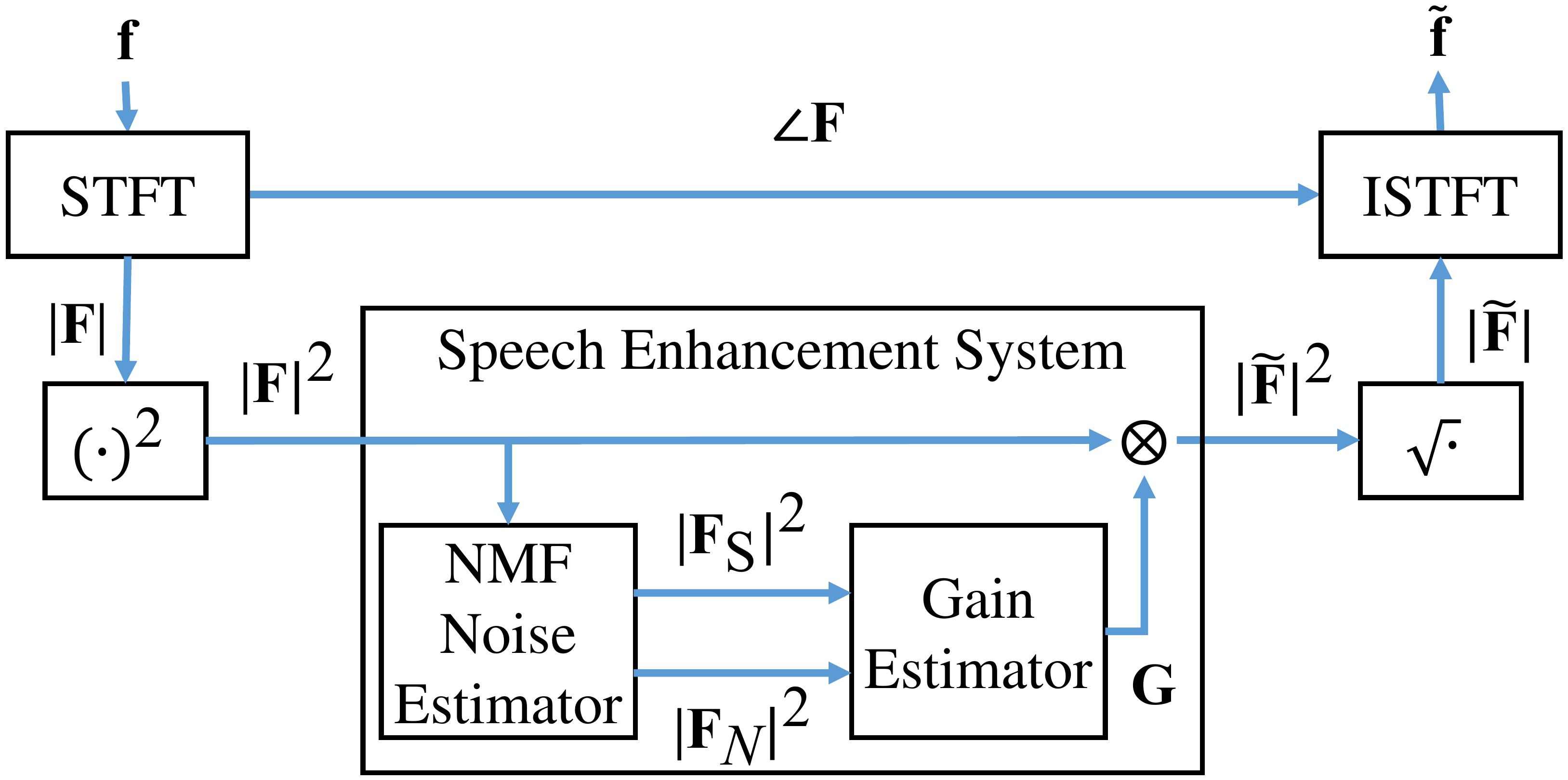}
\caption{The structure for a conventional STFT-based speech enhancement system.}\label{fig:SE}
\end{figure}

The conventional speech enhancement (SE) framework based on STFT-wise spectrogram modification is illustrated in Fig. \ref{fig:SE}. The input real-valued time signal $\textbf{f}$ is first converted to its complex-valued spectrogram $\mathbf{F}$ via STFT. Then the SE system compensates the magnitude part $|\mathbf{F}|$ of the spectrogram while keeps its phase part $\angle \textbf{F}$ unaltered. Finally, the new spectrogram $\tilde{\textbf{F}}=|\tilde{\textbf{F}}|\exp(j\angle\textbf{F})$ with an updated magnitude $|\tilde{\textbf{F}}|$ and the original phase $\angle \textbf{F}$ is converted back to the time domain via inverse STFT (ISTFT) to constitute the enhanced time signal $\tilde{\textbf{f}}$.
%
%

\subsection{NMF noise estimation}
Given a nonnegative matrix $\textbf{V}$, NMF aims to find another two nonnegative matrices $\textbf{W}$ and $\textbf{H}$ such that the difference between $\textbf{V}$ and the product $\textbf{WH}$ is minimized. Here a nonnegative matrix refers to a matrix having nonnegative entries only. In this study, we use the accumulated squared element-wise distance, viz. $d=\sum_{ij}(\textbf{V}_{ij}-(\textbf{WH})_{ij})^2$, as the matrix difference, and applying the multiplication rules as in Eqs. (\ref{eq_nmfW}) and (\ref{eq_nmfH}) iteratively for updating the two matrices $\textbf{W}$ and $\textbf{H}$ can   
decrease the difference $d$. The locally optimal $\textbf{W}$ and $\textbf{H}$ are obtained accordingly when the iteration process converges.

\begin{equation}\label{eq_nmfW}
\textbf{W}\leftarrow\textbf{W}.\times (\textbf{V}\textbf{H}^T)./(\textbf{W}\textbf{H}\textbf{H}^T),
\end{equation}
\begin{equation}\label{eq_nmfH}
\textbf{H}\leftarrow\textbf{H}.\times (\textbf{W}^T\textbf{V})./(\textbf{W}^T\textbf{W}\textbf{H}),
\end{equation}
where ``$.\times$'' and ``$./$'' denote the element-wise multiplication and division, respectively. 

Notably, the matrix $\textbf{W}$ aforementioned is often viewed as a dictionary matrix, which columns can nearly serve as the basis for the column space of the matrix $\textbf{V}$.    In the following, we introduce the noise tracking procedures of an NMF-based SE method that applies to the STFT-derived spectrogram, which in general consists of an off-line phase and an on-line phase.
\subsubsection{Off-line phase}

For the off-line phase that gathers the information of both speech and noise, the power or magnitude spectra of the short-time segments for of the clean speech utterances in the training set are arranged as the columns of a matrix, which serves as the data matrix $\textbf{V}$ for the subsequent NMF processing to produce a noise-free speech spectral basis matrix $\textbf{W}_S$. Besides, the pure noise spectral basis matrix $\textbf{W}_N$ is obtained in the same manner as $\textbf{W}_S$ using the speech-free noise in the training set. 

\subsubsection{On-line phase}
For the on-line phase that deals with a received noise-corrupted speech, the respective power/magnitude spectrogram, denoted by $\textbf{V}_{SN}$ here, is factorized via NMF with the fixed basis matrix $\textbf{W}_{SN}=[\textbf{W}_{S}\:\:\textbf{W}_{N}]$, which is just derived from the off-line phase. Accordingly, $\textbf{V}_{SN}$ can be approximately decomposed into speech part and noise part as follows,
\begin{equation}\label{eq_NMFREC}
\begin{aligned}
\textbf{V}_{SN}&\approx\textbf{W}_{SN}\textbf{H}_{SN}=[\textbf{W}_{S}\:\:\textbf{W}_{N}]\begin{bmatrix} \textbf{H}_{S} \\ \textbf{H}_{N} \end{bmatrix}\\
&=\textbf{W}_S\textbf{H}_S+\textbf{W}_N\textbf{H}_N.
\end{aligned}
\end{equation}
where $\textbf{W}_S\textbf{H}_S$ and $\textbf{W}_N\mathbf{H}_N$ respectively approximate the  speech and noise components in $\textbf{V}_{SN}$.
\subsection{Signal gain estimation}
The gain that applies to the noise-corrupted spectrogram is represented as follows,
\begin{equation}\label{eq_wienerfilter}
\textbf{G}=\textbf{W}_S\textbf{H}_S./(\textbf{W}_S\textbf{H}_S+\textbf{W}_N\textbf{H}_N).
\end{equation}
It is noteworthy that when the power spectrogram is used for analysis, the gain function $\textbf{G}$ shown in Eq. (\ref{eq_wienerfilter}) behaves as a Wiener filter. Besides, this NMF-based SE method that applies to the STFT-derived spectrogram is denoted by a shorthand notation ``STFT-NMF'' in later discussions for simplicity.
\section{DWPT-NMF SPEECH ENHANCEMENT SYSTEM}\label{sec:DWPTNMFSE}
\begin{figure}[!th]
\centering
\includegraphics[width=.97\columnwidth] {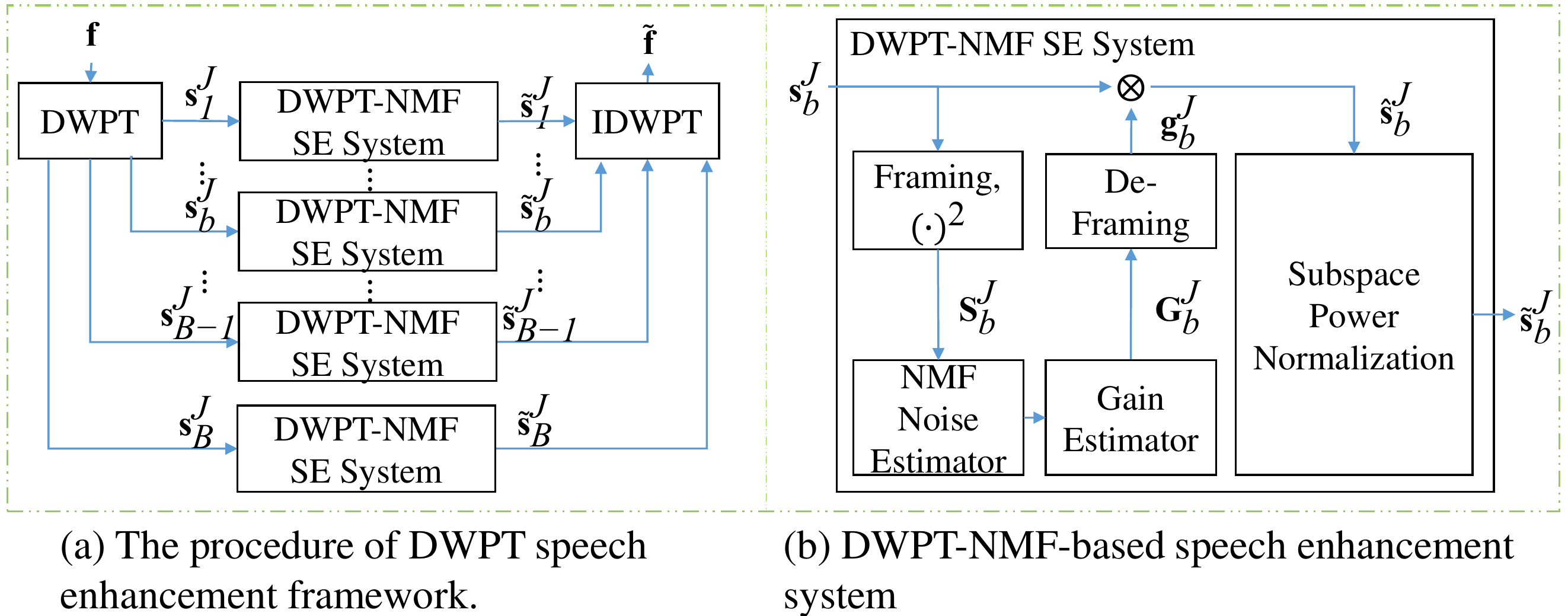}
\caption{The structure for proposed DWPT-NMF speech enhancement system.}\label{fig:DWPTSE}
\end{figure}

In this study, we propose a novel SE method that adopts NMF-wise compensation directly on the time signals, while these time signals are in fact the DWPT (filtered and down-sampled) subband outputs for the original time signal. The block diagram about the overall framework of this SE method is depicted in Fig. \ref{fig:DWPTSE}(a), while the detailed enhancement procedures for each DWPT subband signal is plotted in Fig. \ref{fig:DWPTSE} (b). For simplicity, we use a shorthand notation ``DWPT-NNF'' to stand for this newly proposed SE method. 

As shown in Fig. \ref{fig:DWPTSE}(a), the DWPT is first applied to a noise-corrupted signal $\textbf{f}$ to produce a set of subband signals $\{\textbf{s}_b^J\}$. Then all of the subband signals are individually enhanced via NMF-wise compensation. Finally, these updated subband signals are joined together via the inverse DWPT and the ultimate enhanced signal $\tilde{\textbf{f}}$ is accordingly constructed.

The SE procedures for each subband signal as depicted in Fig. \ref{fig:DWPTSE}(b) are clarified in the following.  The subband signal $\mathbf{s}_b^J$ is first segmented into overlapped frames without further windowing (equivalent to using a rectangular window), and these frames are then arranged in sequence to be the columns of a matrix. Each element of the matrix is further squared in order to produce a nonnegative matrix, denoted by $\textbf{S}_b^J$, for the subsequent NMF processing. 
\subsection{NMF-based noise tracking and gain estimation}
Similar with the NMF noise estimation procedures stated in Section 3.1,  two phases are adopted here. At the off-line phase, the nonnegative matrices $\textbf{S}_b^J$ for the clean utterances in the training set with respect to a specific subband $b$ are concatenated and then analyzed by NMF to create a speech basis matrix $\textbf{W}_S^b$.  Likewise, the noise basis matrix $\textbf{W}_N^b$ associated with the subband $b$ is  created thereby using the speech-free noise in the training set.  As for the on-line phase, the matrix $\textbf{S}_b^J$ of the DWPT subband signal $\textbf{s}_b^J$ for the input noise-corrupted utterance is NMF-encoded with the fixed basis matrix  $\textbf{W}^b= [\textbf{W}_{S}^b\:\:\textbf{W}_{N}^b]$ such that $\textbf{S}_b^J\approx\textbf{W}^b\textbf{H}^b=\textbf{W}_S^b\textbf{H}_S^b+\textbf{W}_N^b\textbf{H}_N^b$, where $\textbf{H}^b$ denotes the NMF encoding matrix, and $\textbf{H}_S^b$ and $\textbf{H}_S^b$ are respectively the speech and noise partitions of $\textbf{H}^b$.
Furthermore, the gain estimation as for subband $b$ is achieved by 
\begin{equation}\label{eq_gain}
\textbf{G}_b^J=\sqrt{(\textbf{W}_S^b\textbf{H}_S^b)./(\textbf{W}_S^b\textbf{H}_S^b+\textbf{W}_N^b\textbf{H}_N^b)},
\end{equation}
where ``$\sqrt{.}$'' denotes an element-wise square root operation. 

Finally,  the overlap-add process as the de-framing scheme is applied to $\textbf{G}_b^J$ in Eq. (\ref{eq_gain}) to obtain a gain sequence $\textbf{g}_b^J$ that has the same size as the original subband signal $\textbf{s}_b^J$.  Therefore,  $\textbf{s}_b^J$ is modulated with $\textbf{g}_b^J$ to produce a new subband signal as
\begin{equation}\label{eq_subsignal}
\hat{\textbf{s}}_b^J=\textbf{s}_b^J.\times\textbf{g}_b^J.
\end{equation}
Notably, since de-framing is processed on the gain estimate $\textbf{G}_b^J$,  the possible distortion caused by STFT does not occur here.  
\subsection{Subband power normalization}
In order to compensate for the noise effect, a power normalization procedure is applied to the enhanced subband signal $\hat{\textbf{s}}_b^J$ in Eq. (\ref{eq_subsignal}).  At the off-line phase we concatenate the DWPT subband-$b$ signals associated with all the clean speech utterances in the training set as a clean sample set, and then we calculate the root mean square (rms) value for this sample set, denoted by $\sigma_{b,c}$.  At the on-line phase,  the rms value of $\hat{\textbf{s}}_b^J$ is also computed and denoted by $\hat{\sigma}_b$, and then we obtain the power normalized subband signal by $\tilde{\textbf{s}}_b^J=\frac{\sigma_{b,c}}{\hat{\sigma}_b}\hat{\textbf{s}}_b^J$. As a result, the power of the ultimately enhanced subband signal $\tilde{\textbf{s}}_b^J$ is always equal to $\sigma_{b,c}^2$,  regardless of different noise-corrupted signals to be enhanced.\vspace{-0.2cm}
\def\FIGURESCALE{0.33}
\def\TABLESCALE{0.07}
\section{EXPERIMENTS}\label{sec:experiment}
\subsection{Experimental setup}
Evaluation experiments are conducted on the Mandarin hearing in noise test (MHINT) database \cite{wong2007development}, which contains 320 utterances pronounced by a Mandarin male speaker and recorded in a clean condition at a sampling rate of 16 kHz. In this study, these utterances were down-sampled to be 8-kHz data.  The averaged length of each utterance is around 3 seconds.  Among these 320 utterances, 10 utterances are selected as the training data and another 50 utterances are the testing data for the speech enhancement task.  In addition, eight types of noise: subway, exhibition, car, street, restaurant, babble, airport, and train-station,  from Aurora-2 database are artificially added to the clean testing utterances to generate the noise-corrupted counterparts at six signal-to-noise ratios (SNRs) ranging from 20 dB to -5 dB with an 5-dB interval.  Besides, these four types of noise are also used together to create the noise basis matrices of NMF at the off-line phase of noise tracking. Finally, the number of columns of each speech basis matrix ($\textbf{W}_S$ and $\textbf{W}_S^b$) is set to 40, while that of each noise basis matrix ($\textbf{W}_N$ and $\textbf{W}_N^b$) is set to 160.

For the STFT-NMF method, the applied frame size and frame shift are 256 samples and 80 samples, respectively. For the proposed DWPT-NMF, a 20-sample frame shift is used, while the frame size is varied. In addition, the level of DWPT/IDWPT is set to 3. 
\subsection{Evaluation methods}
The SE scenarios are evaluated by the quality test in terms of the hearing aids speech quality index (HASQI) \cite{kates2010hearing}, and the perceptual test in terms of the hearing aids speech perception index (HASPI) \cite{kates2014hearing}. Notably, HASQI and HASPI, resepctively, were developed to evaluate sound quality and perception for both hearing impaired patients and normal hearing people. It has been confirmed that these two evaluations provide considerably high correlation scores with human quality assessment and perception. The HASQI and HASPI scores are both ranged from $0$ to $1$. Higher scores of HASQI and HASPI correspond to better sound quality and intelligibility, respectively. 
\subsection{Performance evaluation}
We first investigate the effects of STFT/ISTFT and DWPT/IDWPT on SE processes. In this set of experiments, we prepared paired utterances: a clean utterance and its noisy version. Here the noise utterance is artificially generated by contaminating the clean utterance with the restaurant noise at 0 dB SNR. We first test the transformation performance of STFT/ISTFT. Consider an SE system with STFT/ISTFT: For the magnitude part, we assume that the SE system can perfectly restore the clean magnitude given the noisy magnitude; for the phase part, we consider two conditions: (a) using the noisy phase as the restored phase (denoted as STFT/ISTFT(N)); (b) using the clean phase as the restored phase (denoted as STFT/ISTFT(S)); clearly, the condition (b) represents a perfect restoration in the frequency domain. In addition to STFT/ISTFT, we test the transformation performance of DWPT/IDWPT. Since we assume that the SE system with STFT/ISTFT(S) can perfectly restore both spectrogram and phase parts, for a fair comparison, we assume that the SE system with DWPT/IDWPT can perfectly restore signals in each subband. Table \ref{tab:perfect} lists the results of STFT/ISTFT(N), STFT/ISTFT(S), and DWPT/IDWPT on three evaluations metrics: HASQI, HASPI and mean-squared error (MSE).

From table \ref{tab:perfect}, all three systems have the same the HASPI score, suggesting that the phase part has negligible effect on the intelligibility. However, STFT/ISTFT(N) has lower HASQI score than STFT/ISTFT(S), suggesting that the inaccurate phase information may seriously distort the restored signals. Finally when compared to DWPT/IDWPT, STFT/ISTFT(S) gives a higher MSE value and a lower HASQI score. Please note that STFT/ISTFT(S) has perfect spectrogram and phase information; the result thus confirm that STFT/ISTFT could generate distortions on the enhanced signals.

\begin{table}[!tp]\vspace{-0.1cm}
\begin{scriptsize}
\begin{center}
\caption{Transformed efficiency of STFT and DWPT on HASQI, HASPI, and MSE.}\label{tab:perfect}
\begin{tabularx}{\columnwidth}{c|>{\centering}m{0.175\columnwidth}|>{\centering}m{0.17\columnwidth}|>{\centering\arraybackslash}X}
\hline
\hline
\textbf{Indexes}&\textbf{HASQI}&\textbf{HASPI}&\textbf{MSE}\\
\hline
\hline
\textbf{STFT/ISTFT(N)}&0.623&1.000&0.634\\
\hline
\textbf{STFT/ISTFT(S)}&0.991&1.000&0.523\\
\hline
\textbf{DWPT/IDWPT}&1.000&1.000&0.000\\
\hline
\hline
\end{tabularx}
\end{center}
\end{scriptsize}
\end{table}

\begin{figure}[htbp]
\centering
\includegraphics[width=.57\columnwidth] {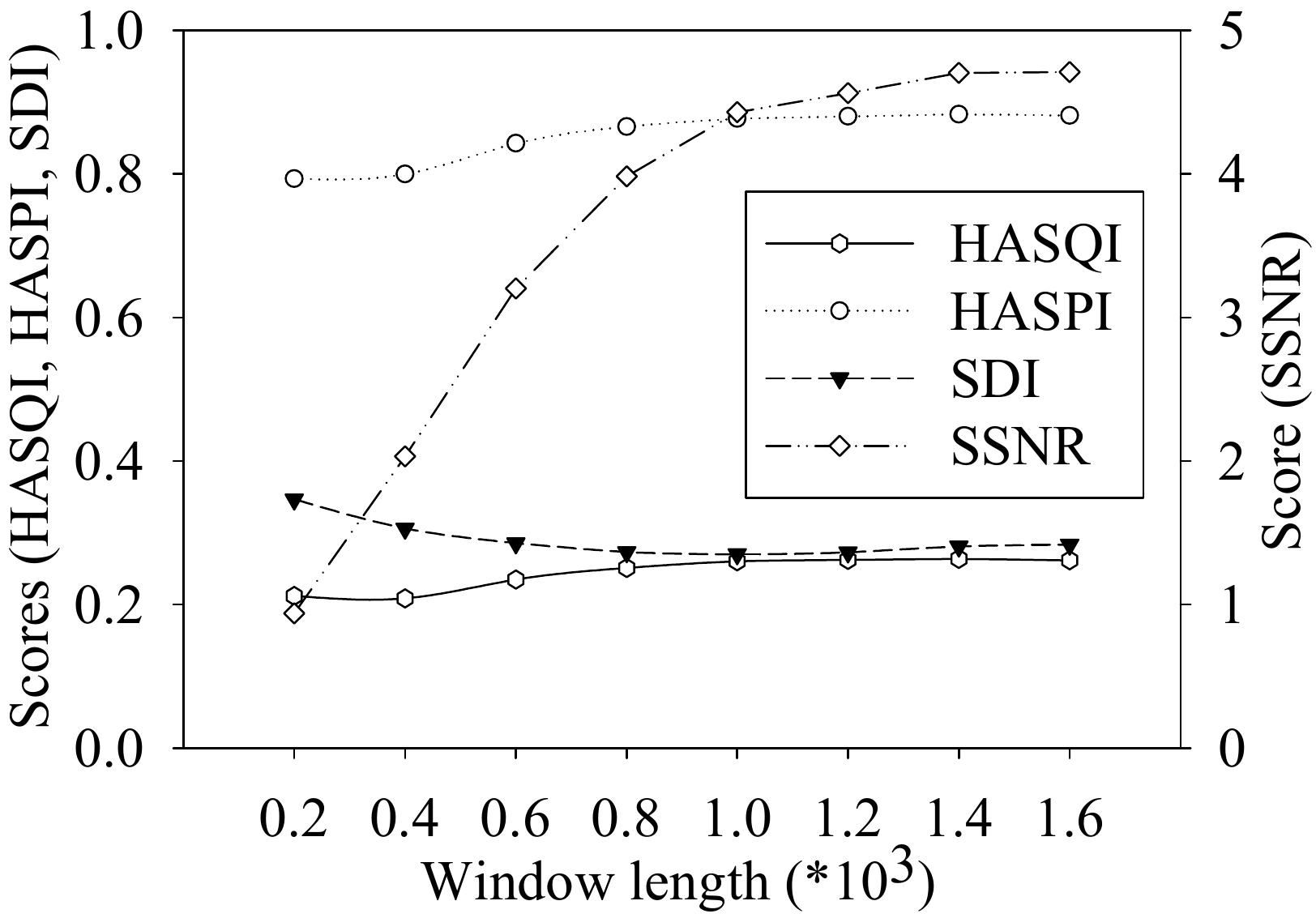}
\caption{The averaged scores of HASQI, HASPI, SDI and SSNR for DWPT-NMF enhanced signals with respect to different framing window lengths}\label{fig:windowchange}
\end{figure}

\begin{figure}[!tp]
\centering
\subfigure[Clean]{
   \includegraphics[width=\FIGURESCALE\columnwidth] {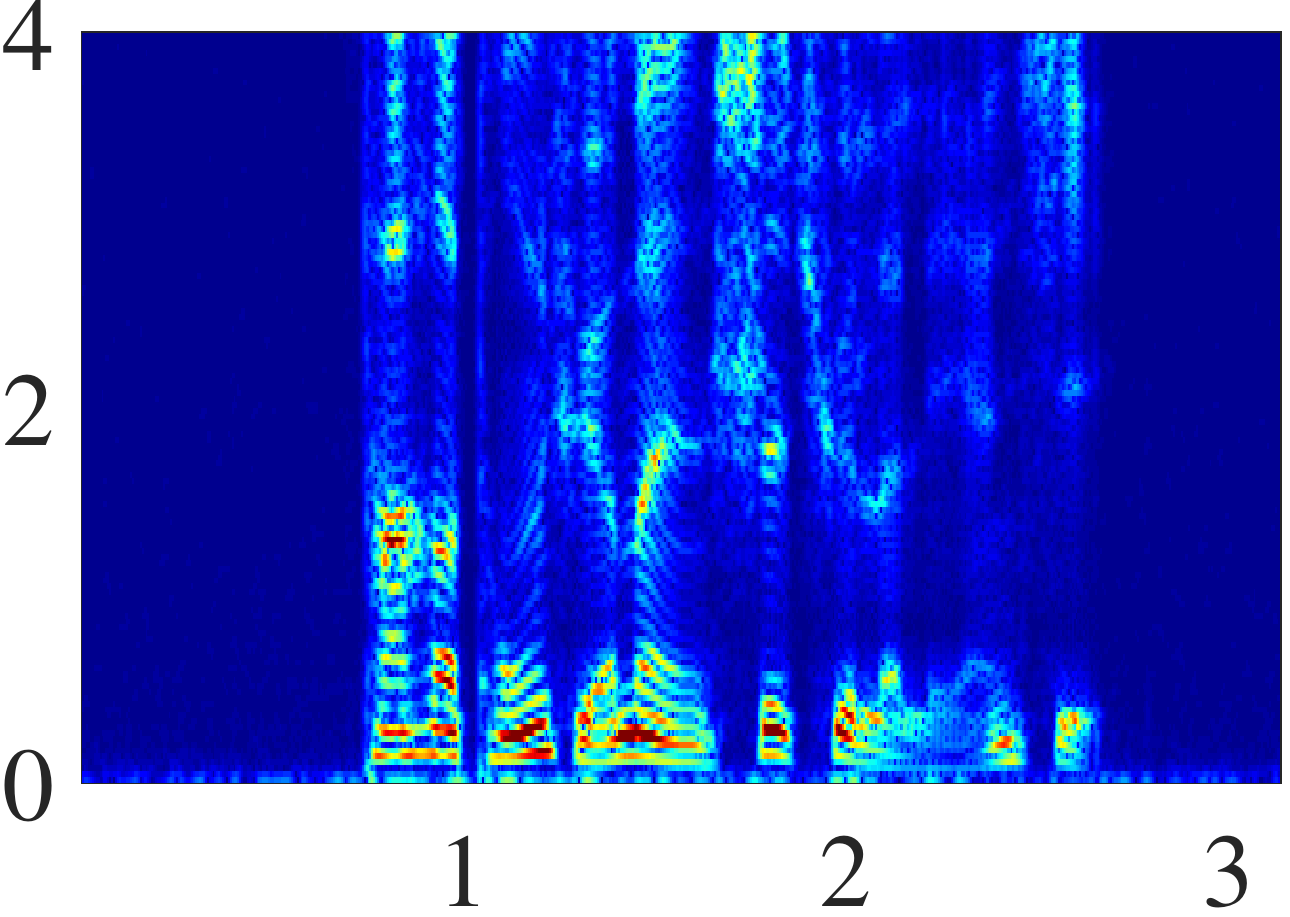}
 }
  \subfigure[Noisy]{
   \includegraphics[width=\FIGURESCALE\columnwidth] {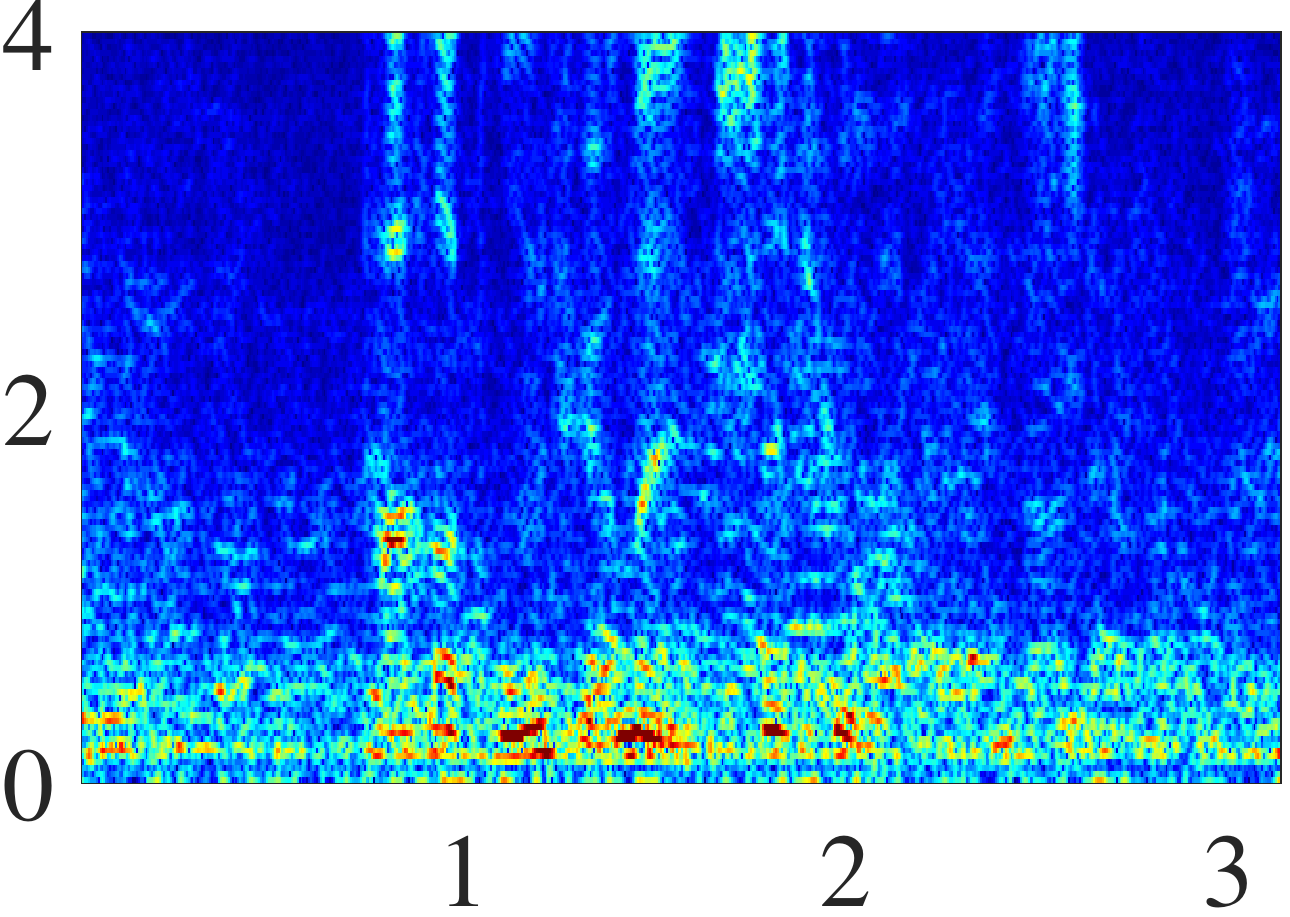}
   }
   \subfigure[Enhanced by STFT-NMF]{
   \includegraphics[width=\FIGURESCALE\columnwidth] {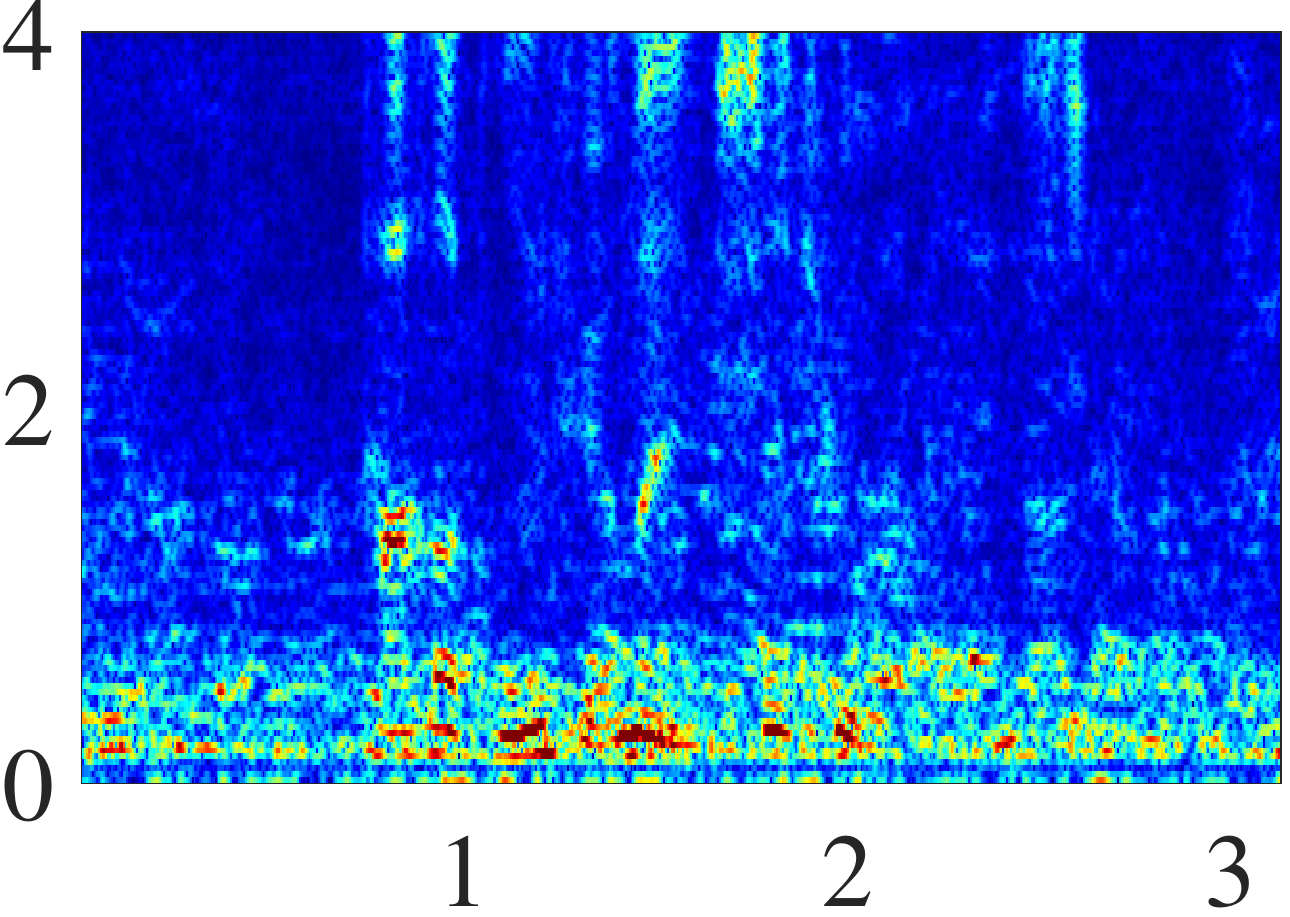}
 }
  \subfigure[Enhanced by DWPT-NMF]{
   \includegraphics[width=\FIGURESCALE\columnwidth] {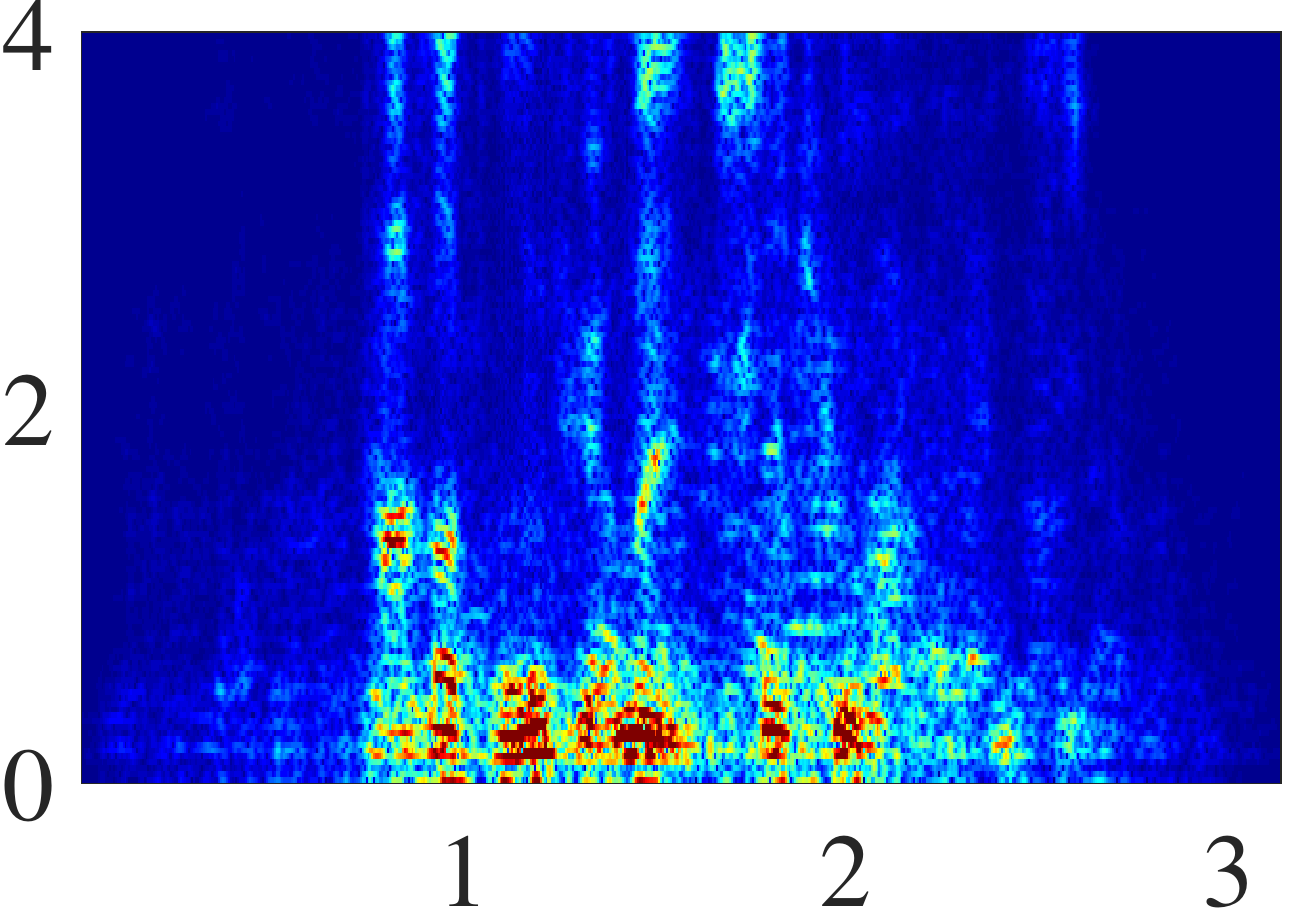}
   }
\caption{The spectrogram of (a) the clean utterance, (b) the noise-corrupted utterance, (c) the noise-corrupted utterance enhanced by STFT-NMF  and (d) the noise-corrupted utterance  enhanced by DWPT-NMF. In all of the four figures the horizontal axis is for time in second and the vertical axis for frequency in kHz. }
\label{fig:spectogram}
\end{figure}
\vspace{-0.2cm}

Second, the effect of different assignments of the frame size in the proposed DWPT-NMF is investigated. The corresponding evaluation is performed on the noise-corrupted data with four noise types and six SNR levels, and the evaluation results in terms of HASQI, HASPI, speech distortion index (SDI) and segmental SNR (SSNR) are shown in Fig. \ref{fig:windowchange}. From the figure, we observe that increasing the length of the framing window from 200 to 1000 always brings about better results in all of the performance evaluation metrics. However, most of the metric scores reach a plateau by further enlarging the frame size greater than 1000. As a result, the frame size is set to 1000 samples in DWPT-NMF for the subsequent experiments.

Third, the quality of the restored spectrogram for an utterance contaminated by babble noise at an SNR of 0 dB is examined visually in Fig. \ref{fig:spectogram}. Comparing these spectrograms, we can see that the conventional STFT-NMF exhibits higher speech distortion as well as more noise residues in the restored spectrogram than the newly proposed DWPT-NMF. Therefore, DWPT-NMF is shown to be superior to STFT-NMF in speech enhancement capability.  

Finally, we compare STFT-NMF and DWPT-NMF in terms of the HASQI and HASPI metric scores   
associated with the enhanced signals. Tables \ref{tab_hasqi} and \ref{tab_haspi} list the corresponding results at six SNR cases while averaged over four noise types. From the two tables, we find that DWPT-NMF gives rise to higher HASQI and HASPI scores than STFT-NMF and the unprocessed baseline in almost all cases, which again reveals that DWPT-NMF is quite effective in improving both the quality and intelligibility of speech signals.

\begin{table}[!tp]
\begin{scriptsize}
\begin{center}
\caption{The HASQI results at six SNR conditions.}\label{tab_hasqi}
\begin{tabularx}{\columnwidth}{>{\centering}X|>{\centering}m{\TABLESCALE\columnwidth}|>{\centering}m{\TABLESCALE\columnwidth}|>{\centering}m{\TABLESCALE\columnwidth}|>{\centering}m{\TABLESCALE\columnwidth}|>{\centering}m{\TABLESCALE\columnwidth}|>{\centering\arraybackslash}m{\TABLESCALE\columnwidth}}
\hline
\hline
\textbf{SNR ($dB$)}&$\mathbf{20}$&$\mathbf{15}$&$\mathbf{10}$&$\mathbf{5}$&$\mathbf{0}$&$\mathbf{-5}$\\
\hline
\hline
\textbf{Baseline}&\textbf{0.452}&0.359&0.261&0.163&0.084&0.036\\
\hline
\textbf{STFT-NMF}&0.418&0.344&0.261&0.175&0.097&0.043\\
\hline
\textbf{DWPT-NMF}&0.447&\textbf{0.394}&\textbf{0.328}&\textbf{0.258}&\textbf{0.178}&\textbf{0.105}\\
\hline
\hline
\end{tabularx}
\end{center}
\end{scriptsize}
\end{table}

\begin{table}[!tp]
\begin{scriptsize}
\begin{center}
\caption{The HASPI results at six SNR conditions.}\label{tab_haspi}
\begin{tabularx}{\columnwidth}{>{\centering}X|>{\centering}m{\TABLESCALE\columnwidth}|>{\centering}m{\TABLESCALE\columnwidth}|>{\centering}m{\TABLESCALE\columnwidth}|>{\centering}m{\TABLESCALE\columnwidth}|>{\centering}m{\TABLESCALE\columnwidth}|>{\centering\arraybackslash}m{\TABLESCALE\columnwidth}}
\hline
\hline
\textbf{SNR ($dB$)}&$\mathbf{20}$&$\mathbf{15}$&$\mathbf{10}$&$\mathbf{5}$&$\mathbf{0}$&$\mathbf{-5}$\\
\hline
\hline
\textbf{Baseline}&0.998&0.994&0.976&0.881&0.541&0.164\\
\hline
\textbf{STFT-NMF}&0.998&0.994&0.977&0.893&0.604&0.212\\
\hline
\textbf{DWPT-NMF}&\textbf{0.999}&\textbf{0.998}&\textbf{0.993}&\textbf{0.977}&\textbf{0.900}&\textbf{0.618}\\
\hline
\hline
\end{tabularx}
\end{center}
\end{scriptsize}
\end{table}\vspace{-0.2cm}
\section{CONCLUSION}\label{sec:concl}
This study proposed a novel speech enhancement framework based on DWPT and NMF. It is shown that DWPT serves as a better choice than the conventional STFT in the preparation of the data for the subsequent NMF-wise enhancement scheme. The proposed DWPT-NMF speech enhancement framework provides the noise-corrupted speech with a significant improvement in both quality and intelligibility. As for the future avenue, we will adopt different noise tracking and gain estimation strategies other than NMF on the DWPT subband signals to see if further improvement can be achieved. Besides, we will test the variants of NMF, such as NMF with a sparse constraint, in the proposed DWPT-NMF.

\bibliographystyle{ieeetr}
\bibliography{\RESOURCEPATH reference}

\end{document}